# Dual Connectivity and Standalone Modes for LTE-U

Yajun Zhao

Baseband Algorithm Department

ZTE Corporation

Beijing, P.R.China

zhao.yajun1@zte.com.cn

Yu-Ngok Ruyue Li, Liujun Hu, Chenchen Zhang

Wireless Advanced Research Department

ZTE Corporation

Shenzhen, P.R.China

{li.ruyue, hu.liujun, zhang.chenchen}@zte.com.cn

*Abstract*—Long-Term Evolution in unlicensed bands (LTE-U) has been considered as an effective way of offloading traffic from licensed bands. This paper discusses the scenarios, requirements and different operation modes of LTE-U. Motivations and benefits of supporting two of the operation modes namely Dual Connectivity(DC) and standalone are discussed. Further, evaluation results of some typical LTE-U scenarios are provided to show the benefits of supporting dual connectivity and standalone modes for LTE-U.

## I. Introduction

The popularity of new wireless devices, e.g. smart-phones, tablets, wearable devices, gives rise to the demand of wireless data traffic. Exponential traffic growth is expected in the future due to the growing number of these devices. The data demand has exposed the issue of potential bandwidth shortage for cellular operators. The cellular industry has tried to address this issue by utilizing the licensed spectrum in more spectrally efficient way. In addition, the opportunistic use of available spectrum in unlicensed bands appears to be a promising and complementary way to address the spectrum crunch [1][2][3][4].

LTE-U (i.e. LTE in unlicensed carrier) is an attractive candidate of 3GPP LTE which has received a lot of attention and strong interest from the industry including operators and vendors. In 3GPP workshop on LTE-U held in June 2014, modes of operation were discussed[5]. LTE-U includes two kinds of operating modes: Licensed-Assisted Access (LAA) mode and Standalone mode. According to the Study Item Description (SID) [6], LAA mode is defined as a licensed-assisted access to unlicensed spectrum in an operator-controlled manner. LAA further includes two candidate modes: Carrier Aggregation (CA) based and Dual Connectivity (DC) based. CA based LAA has been studied and standardized in LTE Rel-13 (only including downlink mechanisms) and Rel-14 (uplink mechanisms). While the former is considered to be the first priority, most of the companies see the value to study Licensed Assisted Dual Connectivity operation as well. Supporting DC or standalone with non-ideal backhaul was mentioned in contribution papers [7][8][9][10][11][12][13].

In LTE, support for wider bandwidth is provided through carrier aggregation (CA) of multiple component carriers(CCs). With CA, users are able to transmit and receive on multiple component carriers simultaneously. Each user can be configured with one downlink and one uplink Primary Component Carriers (PCCs). The remaining carriers are called Secondary Component Carriers (SCCs). To support CA, the carriers seen by the same user have to be either co-located in the same site or non-co-located in different sites with ideal backhaul.

To relax the processing requirement between different carriers, dual connectivity (DC) has been introduced in 3GPP Rel-12 mainly considering heterogeneous network scenarios. A term "Dual Connectivity" is used to refer to operation where a given UE consumes radio resources provided by at least two different network points connected with non-ideal backhaul[14]. In contrast to CA, DC allows the component carriers to be non-co-located in different sites (e.g. macro and small cell) with non-ideal backhaul. The components of DC include Main eNB (MeNB) and Secondary eNB (SeNB) and two of them are connected by non-ideal backhaul. MeNB is considered to provide support of wide coverage and mobility while SeNB (often a small cell) is considered to provide offloading in hotspot area. The component cells of DC on MeNB make up a group called Main Cell Group (MCG) and the component cells on SeNB make up another group called Secondary Cell Group (SCG). Like CA, users by DC are also able to transmit and receive on several component carriers simultaneously. The MAC/RLC module of different component carriers can be non-co-located so that the cells in MCG and SCG can perform scheduling separately.

The scenarios of LTE-U are similar to scenarios of LTE small cell in licensed carrier and Wi-Fi in unlicensed carrier. The scenarios of the latter two have been well studied in many papers [15][16][17]. The scenarios of LTE-U have been studied in some papers [1][2][3] but most of them focus on CA based LTE-U operation mode. The deployment requirements and use cases of LTE-U have not been studied deeply especially for the DC and standalone modes of LTE-U.

In this paper, we will investigate the scenarios of LTE-U and focus on the discussion on DC and standalone operation modes of LTE-U. Motivations and benefits of supporting DC and standalone modes are identified for these scenarios. The remaining part of this paper is structured as follows. Section II presents the requirements, scenarios and operation modes of LTE-U. Section III discusses the motivations and benefits of DC and standalone modes for LTE-U from the perspective of network deployment and UE capability. Evaluation results on DC/standalone operation modes are shown and compared against CA mode in Section IV. Finally, we conclude and give our future work plan in Section V.

## II. REQUIREMENTS, SCENARIOS AND OPERATION MODES

In this section, a brief overview of LTE-U requirements, scenarios and operation modes are provided.

### A. Requirements of LTE in unlicensed carrier

As the existing devices in unlicensed carrier (e.g. Wi-Fi and Bluetooth), LTE-U should follow some rules of unlicensed carrier primarily. Among of them, the most fundamental rule is to share spectrum fairly. There are two major ways for specifying or defining the fairness including:

- Different competitors have the same opportunity to use an unlicensed band.
- Or, Different competitors can obtain a same throughput by using an unlicensed band.

At least, to allow friendly coexistence with Wi-Fi systems or other LTE-U networks deployed by other operators, a listen-before-talk scheme should be introduced. In the listen-before-talk scheme, a device listens to the channel for a period of time and if no ongoing transmission is observed, the device can start its transmission.

Currently, Wi-Fi access nodes enjoy a cost advantage compared to LTE small cells. However, Wi-Fi system is basically designed for local access type services and there is not core network or other central entities for coordination. So Wi-Fi is a distributed system and not suitable for a wide and continual coverage. While, LTE is a type of cellular system and primarily designed for wide and continual coverage. The characteristics of LTE include owning core network, supporting handover, supporting inter-node coordination, etc. Compared to Wi-Fi system, there are several advantages in deploying LTE including [1] better spectral efficiency, better QoS management and control and better mobility support.

It is desirable to maintain most of these original advantages of LTE when LTE is used in unlicensed carrier. Otherwise, the motivation of deploying LTE in unlicensed bands is low as it is considered it is lower cost to deploy the current Wifi system.

### B. LTE-U scenarios and operating modes in unlicesed carrier

The use cases of LAA are considered mostly in heterogeneous network. Fig. 1 illustrates LAA deployment scenarios in heterogeneous network. There are co-located and non-co-located scenarios. When the licensed LTE carrier and unlicensed carrier are co-located in a small cell, CA can be used to achieve LAA.

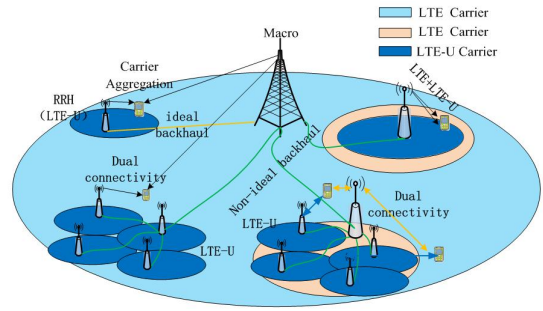

Fig. 1. LAA scenarios

Three types of candidate operating mode of LTE-U have been studied, i.e. CA, DC and Standalone. To meet the fundamental principles of LTE-U design, these three operation modes should have following main common characteristics,

- Follow the fundamental rules of above mentioned, e.g.
    - reusing the LTE core network, i.e. a common core network with LTE in licensed carrier, and
    - supporting inter-cell coordination and handover at least.
- And other principle of operating LTE-U: Try to use unlicensed carrier as much as possible for
    - reducing the cost for renting licensed carrier, and
    - solving the resource limitation of licensed carrier.

These fundamental principles of LTE-U design can also ensure that these three types of operating mode could not be private standalone deployment to replace operator easily.

#### 1) Carrier Aggregation based

The definition of carrier aggregation based (CA-based) is to aggregate carriers in licensed and unlicensed bands. A user is configured with a primary component carrier (PCC) in licensed band and several secondary component carriers (SCC) in unlicensed band. All control-plane (e.g. BCH, RRC and etc.) are carried on the PCC. User-plane data can be transmitted either on the PCC or SCCs. Among three operation modes, CA has the highest requirements for network deployment, which means the high cost and the complexity of the network deployment. At least the following requirements should be met:

- The unlicensed carrier should be co-located with licensed carrier or should be connected to licensed carrier with ideal backhaul.
- The coverage of the cell with unlicensed carrier should be overlapping with the cell with licensed carrier.
- Fine synchronization should be kept between PCC in licensed band and SCC in unlicensed band.

Typical use cases:

- UEs with lower speed.
- UEs located in the coverage overlap area of two types of cells.

Higher QoS traffic can access to a cell with licensed carrier (i.e., PCell) with has higher reliability. The lower QoS traffic can be offload to SCell with unlicensed carriers controlled by PCell dynamically using CA mode. Therefore, the traffic QoS can be guaranteed by the management of the PCell.

*2) Dual-connectivity based*

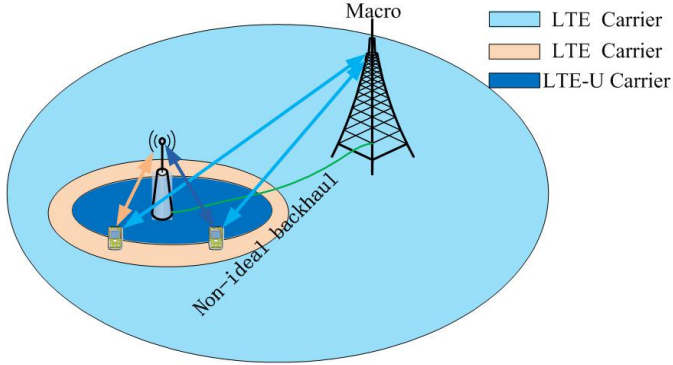

Fig. 2. DC between macro and small cell

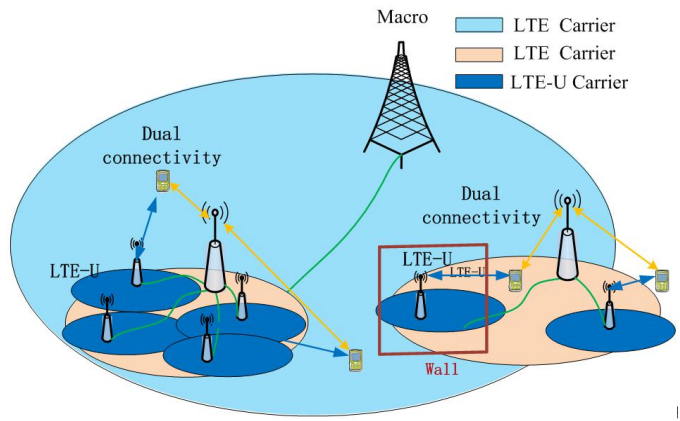

Fig. 3. DC between pico and small cell

As one of most important features in LTE Rel-12, the legacy DC in licensed carrier was studied in depth and has been completed in LTE Rel-12. The application of DC in LTE-U is to perform DC between the cell in licensed band on MeNB and the cell in unlicensed bands on SeNB, and the cell in unlicensed bands on SeNB can be as SCell or PCell of the SCG, i.e. pSCell(as showing in Fig. 2 and Fig. 3). The cell in licensed band on MeNB can be used to support mobility and the transmission of the traffic with higher QoS. At the same time, the cell in unlicensed bands on SeNB can be used to support the transmission of the traffic with lower QoS. Like legacy DC in licensed bands, it is assumed that UEs are not supposed to receive the current broadcasted system information on a cell with unlicensed carrier. At least the following requirement should be met:

- The unlicensed carrier is connected to the licensed carrier with backhaul where the backhaul can be non-ideal

Typical use cases:

- UEs with lower or medium speed.
- UEs located in the coverage overlap area of two types of cells.

The traffic with higher level QoS can be scheduled on cells with licensed carrier and while the traffic with lower level QoS can be scheduled on cells with unlicensed carrier. UE with medium or high speed should also be scheduled with DC mode for supporting better mobility.

*3) Standalone operation based*

The standalone operation mode means that UE can access to a cell with unlicensed carrier independently without the assistance of a cell with licensed carrier. This cell should be connected to a LTE's core network and be able to handover to other cells with unlicensed or licensed carrier for supporting fine mobility and ensuring the continuity of service. Like DC mode, the cells with unlicensed carrier are also covered overlap with the cells with licensed carrier. Among three kinds of model, it is expected that standalone mode is one of the lowest requirements for deployment, which means low cost network deployment with low complexity. It only needs to meet the following requirement:

- The cell on the unlicensed carrier node should be connected to a LTE's core network.

Typical use cases:

- UE with lower speed
- Local area access under good coverage of small cell without any backhaul link to other small cells
- UE traffic with lower level QoS

Obviously, both CA and DC can't be used in this deployment scenario because it can has no backhaul link connected to other nodes with licensed carrier. On the contrary, as mentioned above, standalone mode can work in the deployment scenarios for CA and/or DC. A UE with higher level QoS traffic can also be scheduled as Standalone mode if the candidate unlicensed carriers are in lower load. In this case, LTE-U have more opportunity to use the unlicensed carriers to guarantee the higher level QoS of traffic perfectly. However, when the candidate unlicensed carriers are in high load, the UE has to be scheduled in licensed carrier and can't utilize unlicensed carrier as CA opportunistically. Because, unlike CA mode, the Standalone has no PCell with unlicensed carrier which can operate in licensed spectrum to deliver information with higher priority and guarantee Quality of Service.

Based on the above discussion, we can see that these three operation modes are suitable for different types of traffic and need diffident deployment requirements respectively.

III. NETWORK COVERAGE, DEPLOYMENT CONSTRAINTS AND UE CAPABILIY

In this section, we will analyze the motivations of supporting DC and Standalone in the aspects of coverage, UE

capability and consideration on the constraint of available licensed spectrum and reusing the existing Wi-Fi network.

### A. Small cell density considering different coverage in unlicensed carrier

The typical licensed carriers of small cell are in frequency bands of 2.6GHz and 3.5 GHz and the candidate unlicensed carriers are in frequency band of 5 GHz (especially 5.8 GHz). So, the wireless channel characteristics of licensed and unlicensed bands are different. In general, the propagation path loss of the higher band is larger than lower band which means the unlicensed band can only provide smaller coverage than licensed band using the same transmission power. Besides, the penetration loss in indoor scenario is higher for higher band. This makes the coverage difference exist between licensed bands and unlicensed bands especially for indoor scenarios. The model of the propagation loss is described in [20] and the model includes frequency component. Some typical penetration loss values are provided in [21]. The brief description of the typical models and penetration loss can be found in the following tables.

TABLE I. NDOOR HOTSPOT MODEL

| Scenario | Path loss (dB) |
| --- | --- |
| LoS (Indoor) | $PL = 16.9\ log_{10}(d) + 32.8 + 20\ log_{10}(f_c\text{[a]})$ |
| LoS (Outdoor,UMi) | $PL = 22.0\ log_{10}(d) + 28.0 + 20\ log_{10}(f_c\text{[a]})$ $PL = 40\ log_{10}(d_1) + 7.8 - 18\ log_{10}(h'_{BS}) - 18\ log_{10}(h'_{UT}) + 2\ log_{10}(f_c)$ |

[a.] $f_c$ is given in GHz and distance in m

TABLE II. TYPICAL PENETRATION LOSS OF INDOOR

| Carrier frequency | 2 GHz | 3.5 GHz | 5.0GHz |
| --- | --- | --- | --- |
| Penetration | $20dB+0.5d_{in}$ ($d_{in}$: independent uniform random value between [ 0, min(25,d) ] for each link) | $23dB+0.5d_{in}$ ($d_{in}$: independent uniform random value between [ 0, min(25,UE-to-eNB distance) ] for each link) | $27dB+0.5d_{in}$ ($d_{in}$: independent uniform random value between [ 0, min(25,UE-to-eNB distance) ] for each link) |

From the above tables, we can see the variation on pathloss with frequency, i.e. the pathloss increases with the increase of the frequency.

Some typical simulation results of the coverage difference in indoor scenario without the penetration loss are provided in section IV. From the results of Fig. 6, it can be observed that the big coverage gap between of typical licensed band and of unlicensed band. If the penetration loss is added, the coverage gap becomes even larger.

In addition, higher transmission power is often configured in licensed band than in unlicensed band. The range of the transmission power on a licensed band of small cell can be found in [18] while the transmission power limits of unlicensed band can be found from [19] which are applicable to the system as a whole and in any possible configuration. As we know, the unlicensed carrier is more sensitive to interference. For the purpose of interference coordination, the actual transmission power of unlicensed carrier will probably be lower in most cases.

Due to different channel characteristics and transmit power between unlicensed and licensed bands, there are coverage holes in unlicensed band if cell planning is done according to licensed band. Small cells with unlicensed band need to have denser deployment comparing with the small cell with licensed band to obtain the same coverage. Another way is to deploy more small cell dual mode small cell (i.e. supporting licensed and unlicensed carrier) in denser manner considering full coverage of unlicensed band. However, this will increase the cost of network deployment and create overlapping coverage for licensed band which requires careful interference coordination. In order to reduce the cost and interference, it is desirable to deploy small cells with unlicensed band in different density. To get a similar coverage to licensed carrier, the simulation results of Fig. 6 in Section IV show that the deployment number of nodes with unlicensed carrier is four times than the number of nodes with licensed carrier. These denser nodes with unlicensed carrier should be supported to connect with non-ideal backhaul either to the nodes with licensed band or to core network. With different densities, it is possible that the UE is connected to small cells in different locations. For the former, i.e. connected to the nodes with licensed band, it is beneficial to support DC in this scenario to allow different deployment densities of small cells with licensed bands and unlicensed bands. For the latter, i.e. connected to the core network, it is beneficial to support Standalone in this scenario.

### B. Network deployment constraints

#### 1) Constraint on available licensed spectrum

Some operators may have limited licensed spectrum, e.g. 20MHz. Due to this limitation, the entire licensed spectrum may be used for macro cell to ensure the coverage. If only CA mode of LTE-U is supported, the small cells have to support both licensed bands and unlicensed bands. However, to deploy small cell in the same licensed carrier with macro, it is required to tackle co-channel interference issue between macro and small cell. An alternative is to deploy small cell only in unlicensed carrier connected to macro-cell or connected to core network directly with non-ideal backhaul. For this case, only DC or Standalone mode of LTE-U can be used. The small cells supporting unlicensed carrier only can work with existing macro via DC via non-ideal backhaul or work in Standalone mode. Meanwhile, the small cells supporting unlicensed carrier only without any backhaul have to work in Standalone mode with connection to the core network.

The macro cell is configured with 20MHz licensed carrier and all small cells are configured with unlicensed carrier. DC can be used between licensed carrier of macro and one or more unlicensed carriers of small cells. The licensed spectrum is applied to maximize coverage range and ensure the robustness of the control signalling. Meanwhile, the LTE-U nodes are used to satisfy the demand of hotspots for traffic offloading.

#### 2) Reusing WiFi sites

Currently, many operators already have deployed a large number of Wi-Fi sites which are used for traffic offloading. Adding small cells with licensed carrier requires more careful planning. Therefore, the easiest way for those operators is to upgrade the Wi-Fi APs to support LTE-U only as shown in Fig. 4. Considering non-ideal backhaul of current Wi-Fi network, it requires DC or Standalone to work when it is upgraded to LTE-U. Compared with Wi-Fi, LTE-U is easier in an operator-controlled manner and mobility can be ensured with DC and Standalone.

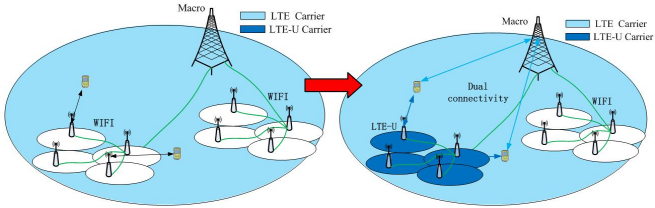

Fig. 4. Reusing Wifi sites for LTE-U upgrade

In this way, the operators can reduce the deployment cost and can deploy LTE-U quickly. On the contrary, if CA based LTE-U is used, a new LTE-U network may be needed with introducing ideal backhaul, planning sites considering interference in licensed carrier etc. It is more difficult to reuse the existing Wi-Fi infrastructure. This means higher network deployment cost and longer time-to-market is required for the CA based solution.

*C. UE carrier capability*

The UE capability of supporting carrier number is limited by number of RF chain, soft buffer, process capability and so on, which mean more expenses.

To support CA, the UE category is usually at least Cat. 4 for commercial UEs in the market. However, if the CA/DC capable UE only supports two carriers (e.g. Cat.4/5 commercial UEs), the UE can only support DC between licensed carrier in macro and licensed carrier in small cell if DC is not supported for LTE-U. This is because one carrier has to be connected to macro for ensuring robustness on mobility. In order to make 2-CC capable UEs benefit from LTE-U, DC has to be supported in LTE-U so that another carrier can be an unlicensed carrier of small cell. Otherwise, this type of UEs could not use unlicensed carrier for traffic offloading in this scenario.

Even for higher cost UEs supporting higher categories with more carriers (e.g. N carriers), to use DC based in licensed carriers, the UEs have to connect to at least two licensed carriers (one licensed carrier from macro and one licensed carrier from small cell) at the same time. If DC is not supported in LTE-U, the UE can only connect N-2 unlicensed carriers using CA with the licensed carrier in the small cell. Considering the current available maximum UE capability is supporting three carriers (i.e. N=3), this means the UE can only connect one unlicensed carrier. This will greatly impose the limitation of the offloading effect of unlicensed carrier. The simulation result shows the performance difference of DC and CA in section IV.

In order to fully utilize the unlicensed carriers, it is desirable to support DC or standalone in LTE-U considering UE capability. With DC, all the UE carriers can be used for LTE-U except the PCell carrier which is connected to macro in licensed band. With standalone mode, all the UE carriers can be used for LTE-U.

IV. EVALUATION RESULTS

In this section, two performance evaluations are carried out for the indoor deployment defined in [21] as part of the 3GPP evaluation scenarios for LTE-U.

- One is the coverage comparison between licensed carrier and unlicensed carrier;
- another is performance evaluation of three LTE-U operation modes, i.e. CA, DC and Standalone, considering limitation of UE carrier capability.

The detailed simulation parameters and other simulation assumptions are also provided.

*A. Coverage comparing between licensed carrier and unlicensed carrier*

In this subsection, we evaluate the performance of coverage in an indoor office (W: 50m, L: 120m) environment composed of fixed cell layout based on the number of small cells as shown in Fig. 5. In Fig 5(a) small cells with licensed carrier and small cells with unlicensed carrier are co-located. Based on this, 4 and 12 more small cells with unlicensed carriers are added in Fig. 5(b) and (c) respectively. Path loss and shadowing are modeled, while the penetration is not considered. Simulations are performed for licensed carrier frequency in 2.6GHz and for unlicensed carrier frequency in 5.8GHz, but the general conclusion can be extended for any other licensed or unlicensed band easily. TABLE III. summarizes the main simulation parameters.

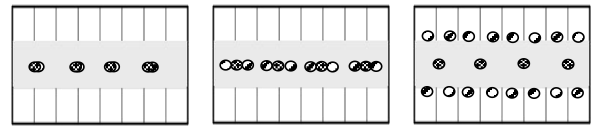

Fig. 5. Deployment with ratio of small cells with licensed carrier to small cells with unlicensed carrier. from left to right (a) 4:4 (b) 4:8 (c) 4:16

TABLE III. SIMULATION PARAMETERS

| Parameters | Value |
| --- | --- |
| Channel Model | ITU InH |
| Antenna pattern | 2D Omni-directional |
| Total BS TX power | 24dBm |

If the two types of small cells with licensed carrier and with unlicensed are deployed in the same density co-located as shown in Fig. 5 (a), i.e. the same number by four nodes are deployed. Fig. 6 shows that the gap of coverage is more than 10dB. Since the cell location for indoor could be determined to cover the whole area with the equal coverage of the two types of cell, location and density of these small cells should be carefully considered. We then try to deploy denser small cells with unlicensed carrier like Fig. 5 (b) and (c). The results of Fig. 6 show that the coverage gap is about 7dB for double number of small cells with unlicensed carrier (i.e. deployed like Fig. 5 (b)). When the number of small cells with unlicensed carrier is increased to 16 and the small cells are deployed like Fig. 5 (c), their coverage is close to the small cells with licensed carrier.

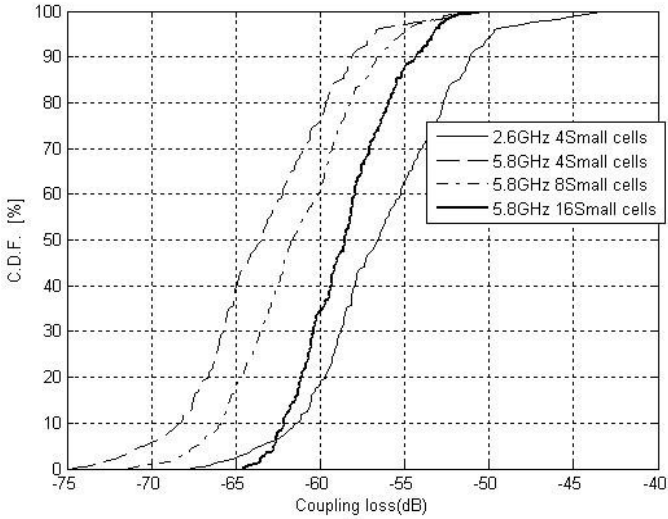

Fig. 6. Coverage comparison between licensed carrier and unlicensed carrier

From the above results, it can be observed that the nodes of the small cell with unlicensed carrier have to be deployed with higher density in order to achieve the same coverage as licensed carrier. This means these two types of small cells with unlicensed and licensed carriers are better to be deployed in non-colocated manner. Typically inter-site nodes has non-ideal backhaul or even no backhaul in between. So, for such scenarios, LTE-U should be operated in DC or Standalone mode instead of CA.

*B. Performance comparing between two LTE-U operation modes of DC/Standalone Vs CA considering limitation of UE capability*

In this subsection, we evaluate the performance of DC mode and standalone mode against CA mode considering limitation of UE carrier capability. The layout of small cell nodes is the same as Fig. 5 (a) underlaying a macro-cell layer for coverage. The number of small cell nodes is four and each node is configured with two types of carriers. There are one licensed carrier with bandwidth of 10MHz and multiple unlicensed carriers with bandwidth of 20MHz. Meanwhile, the carrier capability of UEs is three.

If a UE with three carrier capability working in CA based mode of LTE-U is considered, one of the carriers typically is connected to macro to ensure mobility and coverage. In order to have traffic offloading, it is desirable to connect other carriers to a small cell in proximity. However, if DC mode for LTE-U is not supported, one of them have to be licensed carrier of the small cell in order to use DC between macro and small cell in licensed band. Therefore, only one carrier is left for unlicensed carrier. The licensed carrier on small cell node and one unlicensed carrier on the same node are operated in CA mode of LTE-U. While, for DC based mode of LTE-U, DC between the carrier in licensed band on macro-cell (i.e. MeNB) and the carrier in unlicensed band in small cell node (i.e. SeNB) is possible. It means that only one licensed carrier is required for the UE and the rest of the two carriers can be allocated to unlicensed band simultaneously. For UEs in standalone mode, all the unlicensed carriers can be allocated to these UEs. Static UEs in good coverage of small cell e.g. at the cell center doesn't need to have macro for coverage. In this simulation, we consider total 4 carriers in the system and UEs in different modes have different carrier candidates as summarized in the table IV and table V.

TABLE IV. AVAILABLE CARRIERS IN THE SYSTEM

| Carrier 1 | Macro licensed carrier in 2GHz band |
|---|---|
| Carrier 2 | Small cell licensed carrier in 3.5GHz band |
| Carrier 3 | First small cell unlicensed carrier in 5.8GHz band |
| Carrier 4 | Second small cell unlicensed carrier in 5.8GHz band |

TABLE V. CANDIDATE CARRIER FOR UEs IN CA MODE AND DC MODE

|  | UEs in CA mode | UEs in DC mode | UEs in standalone mode |
|---|---|---|---|
| CC1 | Carrier 1 | Carrier 1 |  |
| CC2 | Carrier 2 | Carrier 2/3/4 | Carrier 3 |
| CC3 | Carrier 3/4 | Carrier 2 /3/4 | Carrier 4 |

Further, two carrier allocation methods are used in the simulation: fixed carrier allocation, or flexible carrier allocation. For the fixed mode, uniform allocation is done such that the number of UEs allocated in each unlicensed bands is the same on the average. Once the allocation is done at the initialization, it is fixed throughout the simulation. For flexible mode, carrier selection from the small cell is done by CSI feedback.

Evaluation is done to compare the following two cases.

Case 1 (CA case) - All UEs can support LTE-U in CA mode only

Case 2 (DC/standalone case) - UEs can support LTE-U in CA/DC/standalone modes and select one of the modes according to the loading and CSI feedback.

TABLE VI. PERFORMANCE GAINS OF DS/STANDALONE VS. CA BASED

| Load | Mean user throughput (Mbits/s) [a] | | | |
|---|---|---|---|---|
|  | *Fixed mode* | | *Flexible mode* | |
|  | CA | DC/Standalone | CA | DC/Standalone |
| 2.5 λ | 173 | 208 | 175 | 236 |
| 10 λ | 136 | 155 | 136 | 163 |

[a.] Only the throughput of the carriers on small cell nodes is counted.

TABLE VII. SIMULATION PARAMETERS

| Parameters | Value |
| --- | --- |
| Number of UEs per node | 20 |
| Packed size | 0.5MBytes |
| Traffic model | FTP mode 1 [22] |
| Receiver | MMSE-IRC |

TABLE VI. shows that, due to more available candidates in unlicensed bands, UEs with DC or standalone support are able to use unlicensed carrier more effectively than CA mode in both low load and high load scenarios. Especially for flexible carrier allocation, the gain is more significant as the UEs can dynamically choose the carrier with better CSI for data transmission. The maximum percentage gain of supporting DC/standalone mode is more than 30%.

V. CONCLUSION

In this paper, scenarios of dual connectivity and standalone mode in LTE-U are discussed. Motivations of supporting DC and standalone are identified. Evaluation is done to show the gain of supporting DC/standalone with limitation of UE carrier capability. Considering the aspects of UE capability, different coverage, constraints of available licensed spectrum and reusing existing Wi-Fi infrastructure, it is beneficial to support DC and Standalone for LTE-U. Therefore, we have the following conclusion:

- DC and Standalone are the two important modes in LTE-U which should be supported in the next step of LTE-U standardization.

The following design direction of next step standardization study should be considered when we study the DC and Standalone of LTE-U in further.

- Study the scenarios and requirements of DC and Standalone in LTE-U.

- Identify the differences among CA, DC and Standalone based solutions.

- Re-use mechanism of legacy DC in licensed carrier as much as possible and identify any specification impact to support a simple DC solution in LTE-U.

- The final mechanism should be a common framework for all three operation modes of LTE-U, even for both TDD and FDD.